\documentclass[12pt,a4paper]{article}
\usepackage{graphicx,amssymb}
\usepackage{graphicx, array, epsfig, subfigure}
\usepackage{subfigure}  
\newcommand{\be}{\begin{equation}}
\newcommand{\ee}{\end{equation}}
\newcommand{\bea}{\begin{eqnarray}}
\newcommand{\eea}{\end{eqnarray}}

\catcode`@=12

\begin{document}
\begin{center}
{\bf Mass and Life Time of Heavy Dark Matter Decaying into IceCube PeV 
Neutrinos}\\
\vspace{1cm}
{{\bf Madhurima Pandey$^{a}$} \footnote {email: madhurima.pandey@saha.ac.in},
{\bf Debasish Majumdar$^{a}$} \footnote {email: debasish.majumdar@saha.ac.in}},\\
{\normalsize \it $^a$Astroparticle Physics and Cosmology Division,}  \\
{\normalsize \it Saha Institute of Nuclear Physics, HBNI}  \\
{\normalsize \it 1/AF Bidhannagar, Kolkata 700064, India } \\
\vspace{0.25cm}
{\bf Ashadul Halder$^{b}$} \footnote{email: ashadul.halder@gmail.com},
{\bf Shibaji Banerjee$^{b}$ \footnote{email: shiva@sxccal.edu}}\\
{\normalsize \it $^{b}$Department of Physics, St. Xavier's College,} \\
{\normalsize \it 30, Mother Teresa Sarani, Kolkata - 700016, India}  \\
\vspace{1cm}
\end{center}

\begin{center}
{\bf Abstract}
\end{center}
{\small
Considering that the ultrahigh energy (UHE) upgoing muon neutrino events 
around the PeV energy region observed by the IceCube are due to the decay of 
super heavy dark matter to neutrinos, we constrain the mass of the decaying 
dark matter and its decay lifetime using the IceCube analysis of these 
neutrinos in the PeV region. The theoretical fluxes are computed by adpoting 
the procedure given in the reference \cite{kuz,bere63}, where the DGLAP 
numerical evolutions of QCD cascades as well as electroweak corrections 
are included 
for evolving the decay process of the super heavy dark matter. Our results 
indicate that to explain the IceCube events around PeV region the decaying 
dark matter mass $m_\chi$ would be $ \sim 5 \times 10^{7}$ GeV with the decay 
lifetime $\tau \sim 7 \times 10^{28}$ sec.
}
\newpage
\section{Introduction}
The origins of the ultra high
neutrino events recorded by the IceCube Collaboration so far are not 
very clearly known. Although one multimessenger
observation in terms of coincident gamma ray detection by the collaboration of 
Fermi-LAT, MAGIC, AGILE, HAWC, H.E.S.S. etc.  
and the high energy neutrino at IceCube-170922A point towards a flaring
blazar TXS0506+056 \cite{messenger}. But the source of other UHE neutrino 
events at IceCube is by and large unknown. These also include the track events 
for the neutrinos in and around PeV region. In this work we 
explore an alternative possibility that these neutrinos could hav been 
created by the rare or long leave decay of superheavy dark matter in the 
Universe \cite{univ1,bere2}. 
The superheavy dark matter could be created during a spontaneous symmetry 
breaking in Grand Unified scale and thus they were never in thermal 
equilibrium with the Universe. Thus their production is nonthermal in nature. 
Such particles can also be originated by the process of gravitational creation 
\cite{univ1} in the early Universe. After the discovery of PeV neutrinos at 
Icecube, the hypothesis of their dark matter origin gained lot of interest 
\cite{icecube6}-\cite{icecube21}. Here in this work, we consider the decay of the super heavy dark matter to interpret the neutrinos 
in and around the PeV region recorded by IceCube that includes the best 
fit region for muon neutrino track events given by IceCube in the same regime.

From the 2078 data sample, where 82 events had been detected by the IceCube 
collaboration, they have made a fit with these 82 events assuming these to have come from extragalactic diffuse UHE neutrino flux \cite{pos}. They have 
fit the flux by 
adopting first an unbroken power law spectrum $ \sim E^{-\gamma}$, 
$\gamma$ being the spectral index and $\gamma$ is obtained as 
$\gamma = 2.92_{-0.29}^{+0.33}$. But later analyses yielded a softer spectrum. 
Also from the upgoing muon events for high energy muon neutrino sample starting from an energy around 120 TeV the IceCube collaboration had obtained a 
different power law. The earlier single power law fit ($\gamma \sim 2.9$)  
which the IceCube collaboration referred to as HESE 
(high energy starting events) data differ from the fit of high energy muon 
neutrino data above 120 TeV. In the present work we consider the softer 
spectrum fit (from the track event) for the later data set above 120 TeV 
and argue that this UHE 
neutrino signals could have originated from the rare decay of very massive 
dark matter with mass $\sim$ $10^7$ GeV. The decay of such massive particles 
generally produces QCD cascade. 

The decay of such massive particles much heavier than electroweak scale 
are 
discussed in the literature such as \cite{kuz,bere63,bere2,bere3}, where the 
decays of super heavy particles are considered to proceed via the cascading of 
QCD partons. It is argued that although the QCD coupling is small the parton 
splitting is favoured by colinear parton emission \cite{bere63}. Also 
these decay processes are enhanced by the electroweak radiative corrections 
at the TeV scale and above. Computer codes have been developed for the 
QCD decay cascade process of such decays where use have been made for 
Dokshitzer-Grivov-Lipatov-Altarelli-Parisi (DGLAP) equations 
\cite{dglap1,dglap2}. In order to treat the electroweak radiative corrections, 
evolution equations similar to DGLAP equations are developed valid for a 
spontaneously broken theory. The electroweak cascade experiences 
interactions of SU(3) $\times$ SU(2) $\times$ U(1) and the couplings 
are enhanced \cite{bere63}. As mentioned in the present work, we are considering a super heavy particle with mass $m_{\chi} \le m_{\rm GUT}$ that decays to 
produce $\nu \bar{\nu}$ ($\chi \rightarrow \nu \bar{\nu}$) as the final 
product. Here for $m_{\chi} > > m_W$ the electroweak cascade accompanies 
the usual QCD cascade. 
The numerical evolution of the DGLAP equations and Monte Carlo (MC) 
studies of such cascades yield the spectrum 
of the final product leptons. In the whole process one needs to consider 
two decay channels, one is the hadronic decay channel while the other is the 
leptonic decay channel. As mentioned in the hadronic decay channel, the 
decay proceeds 
through the QCD cascade whereby the decay of dark matter $\chi$ to 
$\bar{q} q$ ($\chi \rightarrow \bar{q} q$) is first 
produced which 
then hadronizes producing eventually the leptons as final decay products. 
The numerical evolution of DGLAP equations can also be used for the case of 
electroweak radiative corrections. These are studied for a spontaneously 
broken theory \cite{cia}. Similar procedure for electroweak 
cascade attributes to the leptonic decay 
channel. In this work, we compute the neutrino spectrum from the decay of 
super heavy dark matter using first the hadronic channel and then extend to the leptonic channel of dark matter decay to calculate the neutrino flux in the 
PeV region considered. We then make a $\chi^2$ fit with the IceCube reults to 
obtain the values of dark matter mass and its decay lifetime.

In Section. 2 we describe the formalism of our work. Section. 3 gives the 
calculations and the results. Finally a summary and conclusions are given in 
Section. 4.
\section{Formalism}
The final neutrino spectrum is written as \cite{kuz} 
\bea
\displaystyle\frac {dN_{\nu}} {dx} &=& 2R \int_{xR}^{1} \displaystyle \frac 
{dy} {y} D^{{\pi}^{\pm}} (y) + 2 \int_{x}^{1} \displaystyle \frac {dz} {z} 
f_{\nu_i} 
\left (\displaystyle\frac {y} {z} \right) D^{{\pi}^{\pm}} (z)\,\, ,
\label{form1}
\eea
where $D^{\pi} (x,s)$ is defined as $D^{\pi} \equiv [D_{q}^{\pi} (x,s) + 
D_{g}^{\pi} (x,s)]$. $R = \displaystyle\frac {1} {1-r}$, where 
$r = (m_\mu/m_\pi)^2 \simeq 0.573$ and the functions $f_{\nu_i} (x)$ are taken 
from the reference \cite{kelner}
\bea
f_{\nu_i} (x) &=& g_{\nu_i} (x) \Theta (x-r) +(h_{\nu_i}^{(1)} (x) + 
h_{\nu_i}^{(2)} (x))\Theta(r-x) \,\, , \nonumber\\
g_{\nu_\mu} (x) &=& \displaystyle\frac {3-2r} {9(1-r)^2} (9x^2 - 6\ln{x} -4x^3 -5)\,\, , \nonumber\\
h_{\nu_\mu}^{(1)} (x) &=& \displaystyle\frac {3-2r} {9(1-r)^2} (9r^2 - 6\ln{r} -
4r^3 -5)\,\, , \nonumber\\
h_{\nu_\mu}^{(2)} (x) &=& \displaystyle\frac {(1+2r)(r-x)} {9r^2} [9(r+x) - 
4(r^2+rx+x^2)]\,\, , \nonumber\\
g_{\nu_e} (x) &=& \displaystyle\frac {2} {3(1-r)^2} [(1-x) (6(1-x)^2 + 
r(5 + 5x - 4x^2)) + 6r\ln{x}]\,\, \nonumber\\
h_{\nu_e}^{(1)} (x) &=& \displaystyle\frac {2} {3(1-r)^2} [(1-r)
(6-7r+11r^2-4r^3) + 6r \ln{r}]\,\, , \nonumber\\
h_{\nu_e}^{(2)} (x) &=& \displaystyle\frac {2(r-x)} {3r^2} (7r^2 - 4r^3 +7xr 
-4xr^2 - 2x^2 - 4x^2r)\,\, .
\label{form2}
\eea
In this work we compute Eq. (\ref{form1}) and obtained the neutrino spectrum 
for several 
values of $m_{\chi}$. We also found that for the chosen 
range of $m_{\chi}$ in this work the contribution to the neutrino spectrum due 
to the leptonic channel is not more than $\sim 10 \%$. Therfore in this work 
we focus on the hadronic channel. 

The isotropic extragalctic neutrino flux from the decay of such a heavy dark 
matter with mass $m_{\chi}$ is given as 
\bea
\displaystyle\frac {d\Phi_{\rm EG}} {dE} (E_\nu) &=& \displaystyle\frac {1} 
{4\pi m_\chi \tau} \int_{0}^{\infty} \displaystyle\frac {\rho_0 c /H_0} 
{\sqrt{\Omega_m (1+z^3) + (1-\Omega_m)}} \displaystyle\frac {dN} {dE} [E(1+z)] 
dz \,\, .
\label{form3}
\eea
In the above equation (Eq. (\ref{form3})), the proper radius of the Hubble 
sphere, which is known as the Hubble radius, is defined as $c/H_0$, where 
$c/H_0 = 1.37 \times 10^{28}\,\, {\rm cm}$. $\rho_0\,\, (= 1.15 \times 10^{-6}$ GeV/cm$^3$) signifies the average cosmological dark matter density at the 
present epoch (redshift $z = 0$), $\Omega_m = 0.316$ is the contribution of the matter density to the energy density of the Universe in units of the 
critical energy density. The quantity $\displaystyle \frac {dN} {dE}$ in 
Eq. (\ref{form3}) describes the neutrino energy spectrum, which is obtained 
from the decay of super heavy dark matter ($\chi$) and this injected neutrino 
spectrum is considered as a function of the neutrino energy at redsift 
$z$, $E(z) = (1+z)E$.

The galactic neutrino flux from similar decay is described by 
\bea
\displaystyle\frac {d\Phi_{\rm G}} {dE} (E_\nu) &=& \displaystyle\frac {1}
{4\pi m_\chi \tau} \int_{V} \displaystyle\frac {\rho_\chi (R[r])} 
{4\pi r^2} \displaystyle\frac {dN} {dE} (E,l,b) dV \,\, ,
\label{form4}
\eea
where $\rho_\chi (R[r])$ is the dark matter density at a distance $R$ from 
the Galactic Centre and $r$ is the distance from the Earth. In our calculation, for the dark matter density we consider the Navarro-Frenk-White (NFW) profile. 
$l$ and $b$ are the galactic coordinates, where $l$ is the line of sight 
distance. $\displaystyle\frac {dN} {dE} (E,l,b)$ defines the neutrino spectrum 
decaying from the super heavy dark matter particle ($\chi$). We consider the 
Milky way halo over which the integration is taken over in our calculation and 
for which $R_{\rm max}$ is chosen as 260 Kpc. 

The total flux is obtained as
\bea
\phi^{\rm th} (E_\nu) &=& \displaystyle\frac {d\Phi_{\rm EG}} {dE} (E_\nu) + 
\displaystyle\frac {d\Phi_{\rm G}} {dE} (E_\nu)\,\, .
\label{form5}
\eea
The total flux is defined as the theoretical flux $\phi^{\rm th} (E_\nu)$ at an 
energy $E_\nu$ of our analyses.

Also it is assumed that due to the oscillations of the neutrinos the 
neutrinos are 
reaching the earth with the flavour ratio 1:1:1. Using Eq. (\ref{form1} - 
\ref{form5}) we compute the total $\nu_\mu$ flux reaching at the IceCube 
detector and fit them with 
the data points from the pink band given by the IceCube analyses (Figure 2 of 
\cite{pos}).

\section{Calculations and Results} 
We have considered the energy region from $\sim 10^5$ GeV to 
$\sim 5 \times 10^6$ GeV for UHE neutrinos for our analyses. As mentioned 
earlier, this region is obtained by the analyses of the IceCube collaboration 
for the UHE upgoing muon neutrino spectrum and shown as a pink band (for 
1$\sigma$ uncertainties) in Figure 2 of reference \cite{pos}. For our analyses 
we have chosen all the three experimental points that are included in the 
pink band and have adopted several other points within the pink band along 
with their 1$\sigma$ spread (the bandwidth of the pink band at the position of 
the chosen band). The chosen data sets for the fit are given in Table 1.

\begin{table}[]
\centering
\caption{The data points (12 in all) used for the $\chi^2$ fit. First three 
points (marked with ``*") are the observed events by IceCube as shown in 
Figure 1. See text for details}
\vskip 2mm
\label{1}
\begin{tabular}{|c|c|c|}
\hline
Energy  & Neutrino Flux ($E_\nu^2 \displaystyle\frac {d\Phi} {dE}$)  & Error 
\\ 
(in GeV) & (in GeV cm$^{-2}$ s$^{-1}$ sr$^{-1}$) & \\ \hline
2.51189e+06$^*$ & 4.16928e-09$^*$ & 8.2726e-09$^*$ \\ \hline
1.19279e+06$^*$	& 5.03649e-09$^*$ & 7.5383e-09$^*$ \\ \hline
2.68960e+05$^*$	& 7.50551e-09$^*$ & 8.1583e-09$^*$ \\ \hline
3.54813e+06 &	5.25248e-09 &	4.1258e-09 \\ \hline
2.30409e+06 &	5.71267e-09 &	4.1600e-09 \\ \hline
1.52889e+06 &	6.21317e-09 &	3.9882e-09 \\ \hline
1.05925e+06 &	6.61712e-09 &	3.7349e-09 \\ \hline
7.18208e+05 & 	7.04733e-09 &	3.9777e-09 \\ \hline
4.46684e+05 &	7.66476e-09 &	3.6478e-09 \\ \hline
2.86954e+05 &	8.16308e-09 &	4.1571e-09 \\ \hline
1.90409e+05 &	8.87827e-09 &	6.2069e-09 \\ \hline
1.43818e+05 &	9.65612e-09 &	6.8856e-09 \\ \hline
\end{tabular}
\end{table}
The pink band as given by the IceCube Collaboration in Ref. \cite{pos} 
along with the three observational points included in the band are reproduced 
in Figure 1.

\begin{figure}[h!]
\centering
\includegraphics[height=6.0 cm, width=10.0 cm,angle=0]{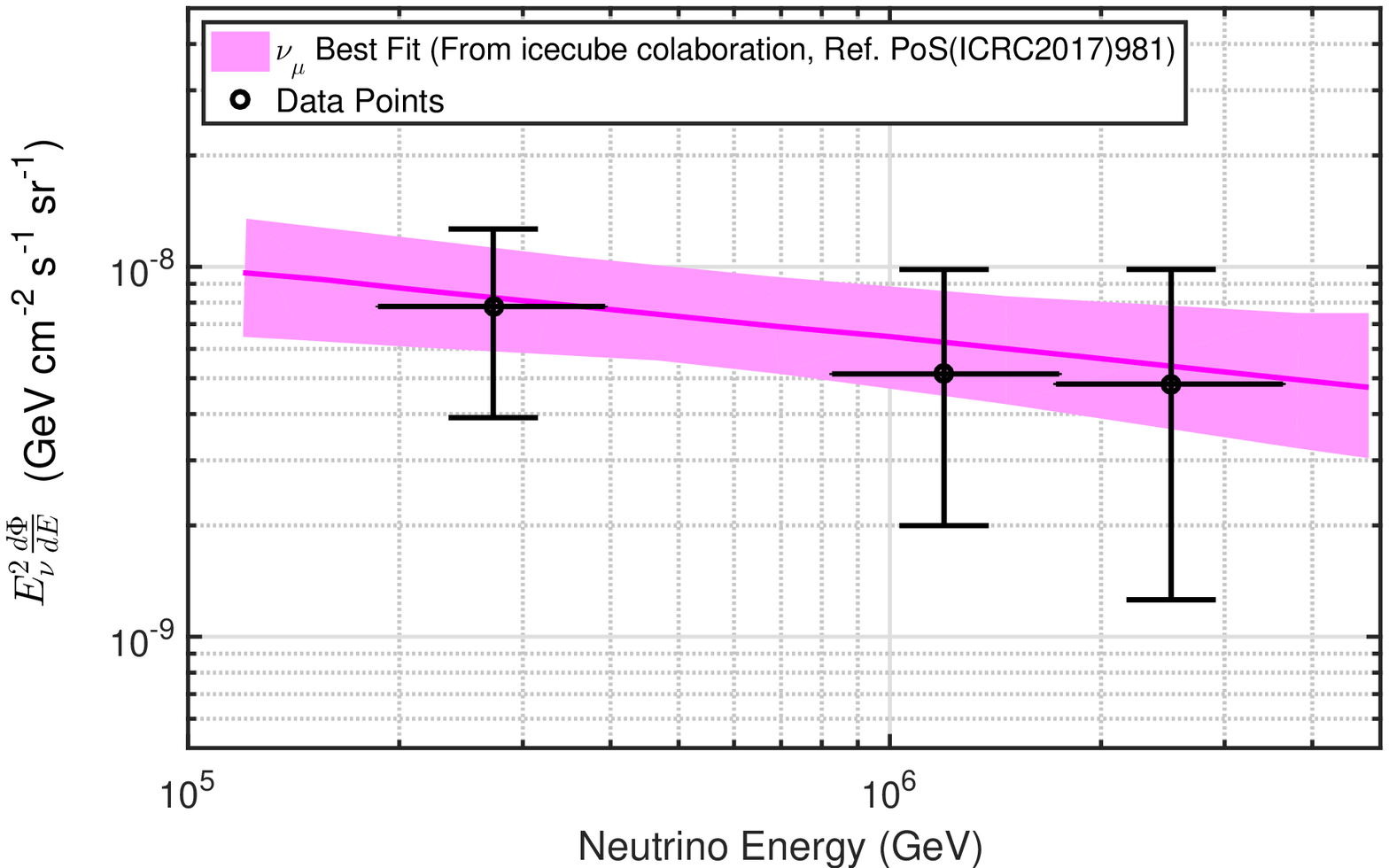}
\caption{The region of IceCube observations adopted for the analysis 
considering dark matter decay (reproduced from \cite{pos}).}
\label{fig1}
\end{figure}

The purpose of this work is to find the best fit value of the mass of the 
super heavy dark matter ($m_{\chi}$) and its decay lifetime ($\tau$) 
that on its decay produce the UHE neutrinos in the region considered. 
The $\chi^2$ for our fit is defined as 
\bea
\chi^2 &=& \sum_{i = 1}^{n} \left (\displaystyle\frac {E_i^2 \phi_i^{\rm th} - 
E_i^2 \phi_i^{\rm Ex}} {(\rm err)_i} \right )^2 \,\, ,
\label{cal1}
\eea
where $n(=12)$ is the number of chosen points (Table 1) and $E_i (i=1,..,n)$ 
are the energies of the chosen points. In Eq. (\ref{cal1}) $\phi_i^{\rm th} 
(E_\nu)$ (and hence $E_i^2 \phi_i^{\rm th} (E_\nu))$, the theoretical flux 
is obtained from Eq. (\ref{form5}) where $E_i^2 \phi_i^{\rm Ex} (E_\nu)$ 
(=$E_\nu^2 \displaystyle\frac {d\Phi} {dE}$) corresponding to experimental 
data are given in Table 1 and $({\rm err})_i$ is the $i^{\rm th}$ chosen 
experimental points (Table 1). 

Chosing a range of $m_{\chi}$ and the decay time $\tau$ we compute the 
$\chi^2$ and one obtained the best fit values of $m_{\chi}$ and $\tau$ by 
minimizing the $\chi^2$ and the minimum $\chi^2$ denoted as 
 $\chi_{\rm min}^2$. The 1$\sigma$, 2$\sigma$ and 3$\sigma$ ranges for 
$m_\chi$ and $\tau$ are also obtained. As the present $\chi^2$ fit is a 
two parameter fit, the 1$\sigma$, 2$\sigma$ and 3$\sigma$ regions are obtained 
by adopting the range of $\chi^2$ to be $\chi_{\rm min}^2 + 2.30, 
\chi_{\rm min}^2 + 4.61, \chi_{\rm min}^2 + 9.21$ respectively. The results are 
shown in Figure 2. 

\begin{figure}[h!]
\centering
\includegraphics[height=6.0 cm, width=8.0 cm,angle=0]{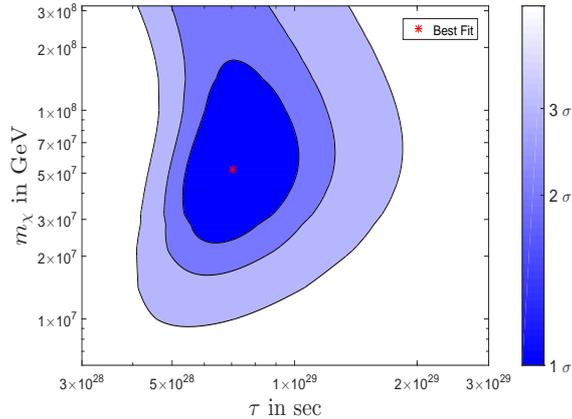}
\caption{$m_\chi - \tau$ (two parameter) $\chi^2$ fit corresponding to the 
$1\sigma, 2\sigma,$ and $3\sigma$ level of confidence. See text for details.}
\label{fig1}
\end{figure}

The best fit value of $m_\chi$ and $\tau$ from our analyses 
are obtained as $m_\chi = 5.2 \times 10^7\,\, {\rm GeV}$, 
$\tau = 7.05 \times 10^{28}\,\, {\rm sec}$. This is denoted by a point in 
Figure 2. The 1$\sigma$, 2$\sigma$ 
and 3$\sigma$ are also shown in Figure 2 by different shades. From this 
analysis it can be said that in case the UHE neutrinos of the chosen energy 
range, adopted from IceCube experimental results, are generated from the 
decay of super heavy dark matter then the mass of such dark matter will be 
$\sim 5 \times 10^7$ GeV undergoing the rare decay with decay lifetime 
$\sim 7 \times 10^{28}\,\, {\rm sec}$. In Figure 3 we show the neutrino flux 
in the PeV region calculated with these best fit values.

\begin{figure}[h!]
\centering
\includegraphics[height=6.0 cm, width=10.0 cm,angle=0]{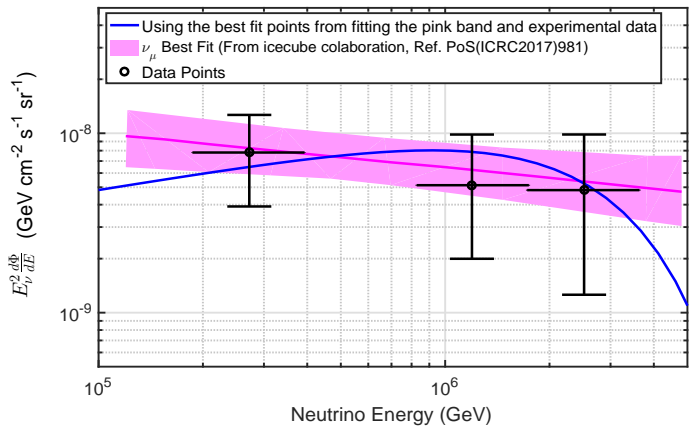}
\caption{The neutrino flux with best fit values of $m_\chi$ and $\tau$. Only 
hadronic channel is considered for dark matter decay.}
\label{fig1}
\end{figure}

In order to explore the effect of the decay of the super 
heavy dark matter via the leptonic channel on the IceCube results we fix 
the mass $m_\chi$ at its 
best fit value obtained above ($m_\chi = 5.2 \times 10^7$ GeV). The neutrino 
spectrum due to the leptonic channel for $m_\chi =5.2  \times 10^7$ GeV is 
evaluated from the formalism given in \cite{kuz,bere63} and the corresponding 
galactic and the extragalactic neutrino fluxes are calculated using the 
relations Eq. 4,5 \cite{kuz}. The total flux is now thereof the sum of the 
fluxes for hadronic channel and the leptonic channel in which the 
decay lifetime $\tau$ is an unknown parameter. We now make a one parameter 
$\chi^2$ analysis (unsing Eq. (\ref{cal1}) for all the set of points given in 
Table 1 and 
obtained the best fit value of the decay lifetime for the dark matter mass 
$m_\chi = 5.2 \times 10^7$ GeV fixed at the best fit value of $m_{\chi}$ 
(Figure 2) when both the hadronic and the 
leptonic channels are considered. We obtain the best fit value $\tau$ to be 
$8.57 \times 10^{27}$ sec. This is further illustrated in Figure 4 where we 
plot the 
variation of $\chi^2$ (one parameter $\tau$) with the fitted parameter $\tau$. 
Thus we see we have a marginal modification of the decay 
lifetime when the leptonic channel effects are included in our analysis. We 
also show in Figure 5 the variation flux with neutrino energies (in the PeV 
region) for the fitted value of $\tau$ as obtained from Figure 4. This 
can be noted in Figure 5
that a kink appears at the tale of the fitted plot. This kink is 
due to the shape of the neutrino spectrum obtained from the leptonic channel 
which showes a sharp rise at higher energy region.

\begin{figure}[h!]
\centering
\includegraphics[height=6.0 cm, width=7.5 cm,angle=0]{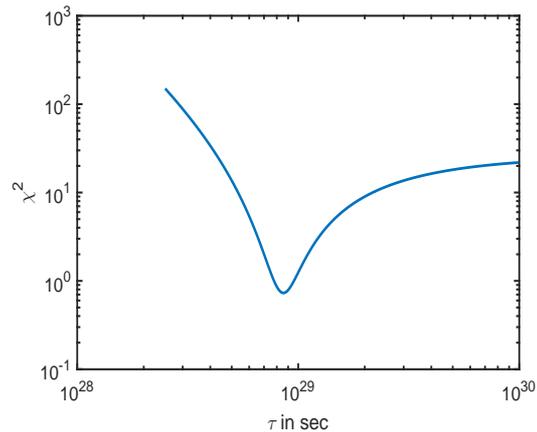}
\caption{Variation of $\chi^2$ with the fitted parameter $\tau$. 
See text for details.}
\label{fig2}
\end{figure}

\begin{figure}[h!]
\centering
\includegraphics[height=6.0 cm, width=10.0 cm,angle=0]{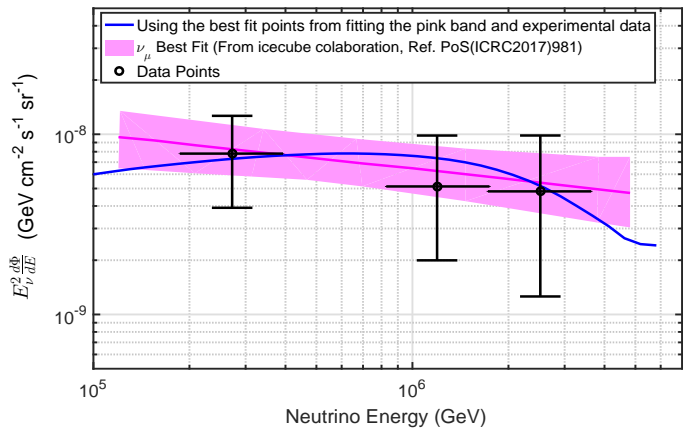}
\caption{The neutrino flux with best fit values of $\tau$ when both 
hadronic and leptonc channel are considered for dark matter decay.}
\label{fig1}
\end{figure}

\section{Summary and Conclusions}
In this work, we have explored the possibility that the UHE neutrino detected 
by IceCube in and around PeV energy region could have originated from 
the decay of super heavy dark matter. Such super heavy dark matter could be 
created at the early Universe during spontaneous symmetry breaking 
at the GUT scale and they decay 
to leptons at the electroweak scale involving the processes of QCD cascade and 
electroweak cascade. The numerical evolution of such process are generally 
done by Monte Carlo methods or by evolving the DGLAP equations treated 
numerically. 
In a recent work M. Kachelriess {\it et al.} \cite{bere63} 
made a MC analysis for such heavy dark matter decay to leptons such as 
$\nu \bar{\nu}$ pair and $e^+ e^-$ pair as well 
as the photons where the recently updated limits of diffuse gamma ray flux has 
also been incorporated. In our present work, we compute the neutrino spectrum 
from the heavy dark matter decay as prescribed in reference and obtained the 
galactic and extragalactic neutrino fluxes from such decays. We then constrain 
the two unknown parametrs namely the heavy dark matter mass ($m_\chi$) and 
its decay lifetime ($\tau$) by making a $\chi^2$ fit of the calculated 
neutrino flux with those 
given from the observed events at IceCube by the IceCube collaboration. 
For this purpose we have chosen the region given by the IceCube collaboration 
corresponding to the upgoing muon neutrinos with energies in and around PeV 
region. The IceCube collaboration analysis designated this region by a 
pink band of 1$\sigma$ width in their published data and plots. We have made a 
$\chi^2$ fit adopting a data set from this region that includes the observed 
points as well as other points choosen from within this pink band. We first 
consider the hadronic channel for $\nu \bar{\nu}$ production from the 
heavy dark matter decay and our $\chi^2$ fit yield the best fit value for 
the parameters $m_\chi, \tau$ as $m_\chi =5.2 \times 10^{7} \,\, {\rm GeV}, 
\tau =7.05 \times 10^{28}\,\, {\rm sec}$. We also 
furnish 1$\sigma, 2\sigma$ and $3\sigma$ C.L. contours in the 
parameter space  $m_\chi, \tau$. With this best fit value of $m_\chi$ we then 
add the contribution of the leptonic channel for this value of $m_\chi$ and then constrain the decay life time $\tau$ by performing the one parameter $\chi^2$ 
analysis. From our studies it appears that in order to explain the nature of 
the neutrino flux for the upgoing muon events in and around PeV region by 
considering the decay of heavy dark matter to neutrinos, the dark matter 
mass should be of the order of $m_\chi \sim 5 \times 10^{7} \,\, {\rm GeV}$ GeV undergoing the rare decays with 
life time $\tau \sim 7.05 \times 10^{28}\,\, {\rm sec}$. Constraining the 
$m_\chi - \tau$ parameter space 
by calculating the spectrum from both the hadronic and the leptonic channel 
and the analysis for the higher energy range that extends upto $10^8$ GeV 
are in progress. It appears that in order to explain the nature of the flux 
towards the $10^8$ GeV regime, the leptonic channel could be very 
important.

{\bf Acknowledgments} : One of the authors (MP) thanks the DST-INSPIRE 
fellowship grant by DST, Govt. of India. Two of the authors (S.B. and A.H.) 
wish to acknowledge the support received from St.Xavier’s College, kolkata 
Central Research Facility. One of the authors (A.H.) also acknowledges 
the University Grant 
Commission (UGC) of the Government of India, for providing financial support, 
in the form of UGC-CSIR NET-JRF.

\end{document}